# Solution-Based Fabrication of High-Performance $K_{0.5}Na_{0.5}NbO_3$ Thin Films for Surface Haptics

Nagamalleswara Rao Alluri,[a,b,*] Longfei Song,[a,b,c,†] Stephanie Girod,[a] Barnik Mandal,[a,c] Juliette Cardoletti,[a,‡] Vid Bobnar,[d] Torsten Granzow,[a,b] Veronika Kovacova,[a,b] Adrian-Marie Philippe,[a,e] Emmanuel Defay,[a,b,c] and Sebastjan Glinsek[a,b]

[a]Smart Materials Unit, Luxembourg Institute of Science and Technology, 41 rue du Brill, L-4422 Belvaux, Luxembourg

[b]Inter-Institutional Research Group Uni.lu–LIST on Ferroic Materials, 41 rue du Brill, L-4422 Belvaux, Luxembourg

[c]Department of Physics and Materials Science, University of Luxembourg, 41 rue du Brill, L-4422 Belvaux, Luxembourg

[d]Department of Condensed Matter Physics, Jožef Stefan Institute, Jamova cesta 39, SI-1000 Ljubljana, Slovenia

[e]Advanced Analyses and Support Unit, Luxembourg Institute of Science and Technology, 41 rue du Brill, L-4422 Belvaux, Luxembourg

---

* Currently at Materials Science and Metallurgical Engineering, Kyungpook National University, 80, Daehak-ro, Buk-gu, Daegu, 41566, South Korea

† Currently at College of Intelligent Systems, Science and Engineering, Harbin Engineering University, Harbin, 150001, China

‡ Currently at Institute of Materials Science, Technische Universität Darmstadt, Peter-Grünberg-Strase 2, Darmstadt, 64287, Germany

**Corresponding authors:** Dr Nagamalleswara Rao Alluri (Dr. Sebastjan Glinsek), Luxembourg Institute of Science and Technology, 41 rue du Brill, L-4422 Belvaux, Luxembourg. e-mail: alluri@knu.ac.kr (sebastjan.glinsek@list.lu).

**Abstract**

$K_{0.5}Na_{0.5}NbO_3$ is among the most promising lead-free piezoelectrics. While its sputtered films match the performance of the champion piezoelectric $Pb(Zr,Ti)O_3$, reproducible processing of high-quality and time-stable solution-deposited $K_{0.5}Na_{0.5}NbO_3$ films remains challenging. Here, we report 1 µm-thick Mn-doped $K_{0.5}Na_{0.5}NbO_3$ films prepared via a chemical solution deposition process, which have perfectly dense microstructure and uniform composition across their thickness. The films exhibit a high transverse piezoelectric coefficient ($e_{31,f}$ = -15.4 C/m$^2$), high dielectric permittivity ($\varepsilon_r \approx 920$), low dielectric losses ($\tan\delta = 0.05$) and can withstand electric fields up to at least 1 MV/cm. Their functional properties show excellent stability over time, and the synthesis process is reproducible. Furthermore, a surface acoustic haptic device based on $K_{0.5}Na_{0.5}NbO_3$ thin-film actuators is demonstrated. The results validate the high potential of Mn-doped $K_{0.5}Na_{0.5}NbO_3$ films to replace lead-based $Pb(Zr,Ti)O_3$ films in piezoelectric applications.

## 1. Introduction

Piezoelectric thin films have been intensively researched and can be applied in different devices, including RF filters, fingerprint sensors, gyro sensors, haptics, and energy harvesters.[1–3] Popular materials for these applications are Sc-doped AlN, $LiNbO_3$, $Pb(Mg_{1/3}Nb_{2/3})O_3$-$PbTiO_3$ (PMN-PT), and $Pb(Zr,Ti)O_3$ (PZT). For sensing and actuator applications, lead-based materials are still considered the champions, with the highest piezoelectric coefficients $e_{31,f}$ ~ -17 C/m$^2$ for PZT.[3] Recently, PZT thin films have proved very effective for texture-rendering haptic applications.[2,4,5] But lead-based piezoelectrics are and will be faced with legislative limitations due to their health and environmental hazards.[6]

Research on lead-free oxide piezoelectric ceramics has therefore been flourishing, with emerging materials like $(K,Na)NbO_3$ (KNN), $(Bi_{1/2}Na_{1/2})TiO_3$ (BNT), and $(Ba,Ca)(Zr,Ti)O_3$ (BCZT).[7–9] Currently, the general understanding is that a single material cannot replace PZT, and the alternatives are application-specific.[10] In thin film form, the most promising material is KNN, but its properties depend strongly on the deposition method. Shibata et al.[11] showed the way to achieve high-quality 3-μm thick KNN films by radio-frequency magnetron sputtering, with a transverse piezoelectric coefficient $e_{31,f}$ of -14.4 C/m$^2$. However, achieving stable piezoelectric and ferroelectric properties of KNN films with low leakage obtained by the cost-effective chemical solution deposition (CSD) process remains challenging.[12–14] Different strategies have been developed to improve CSD-derived films, including doping, optimization of solution chemistry, deposition conditions, and control of film orientation.[7,12–19] Doping KNN with Mn is especially effective in preparing films with decreased leakage current, improved electrical breakdown strength, and good ferroelectric properties.[14] Wang et al.[19] obtained films with low leakage current of $3 \cdot 10^{-6}$ A/cm$^2$ at 50 kV/cm, i.e., 4 orders of magnitude lower compared to pure KNN, by doping with 2 mol.% of Mn. They attributed the enhanced properties to a decreased concentration of holes caused by the increased valence-state of $Mn^{2+}$ to $Mn^{4+}$ incorporated on the perovskite B-sites. Sung et al.[12] achieved a piezoelectric coefficient $e_{31,\mathrm{f}}$ and current density of -8.5 C/m$^2$ and $10^{-7}$ A/cm$^2$ at 100 kV/cm, respectively, in 0.5 mol.% Mn-KNN films. Matsuda et al.[16] reported remanent polarization ($P_\mathrm{r}$), coercive electric field ($E_\mathrm{c}$), and leakage current density of 1 mol.% Mn-KNN films of 14 μC/cm$^2$, 97.5 kV/cm, and $10^{-7}$ A/cm$^2$ at 100 kV/cm, respectively. Kovacova et al.[15] reported 0.5 mol.% Mn-KNN films with leakage current density of $2.8 \cdot 10^{-8}$ A/cm$^2$ at 600 kV/cm, with $P_\mathrm{r}$ and $E_\mathrm{c}$ values of 5.7 μC/cm$^2$ and 50 kV/cm, respectively. To date, CSD-derived KNN films with piezoelectric properties comparable to those of PZT films have not been reported. This is most likely due to the delicate balance between processing parameters, defect formation, and resulting film performance.

In this paper, we report 1 μm-thick CSD-derived 1 mol.% Mn-doped KNN films. The films are single-phase perovskites at X-ray diffraction (XRD) level. They have remanent polarization $P_\mathrm{r}$ and coercive field $E_\mathrm{c}$ values of 10 μC/cm$^2$ and 40 kV/cm, respectively, while the piezoelectric coefficient $e_{31,\mathrm{f}}$ varies between -12.6 and -15.4 C/m$^2$ and is comparable to PZT thin films. Leakage current contribution remains low even at high electric fields (±1 MV/cm). The ferroelectric properties of the films are reproducible and stable with time for at least 7 months. We demonstrate a surface haptic device by inducing an acoustic wave in an Si substrate via two deposited thin-film actuators. The device shows out-of-plane displacements beyond 1 μm at power consumption of 75 mW. This study shows that CSD-based KNN films stand as a credible alternative to replace PZT films in thin-film piezoelectric applications.

## 2. Materials and Methods

### *2.1 Synthesis of Solutions*

50 mL of a 1 mol.% Mn-doped potassium sodium niobate solution ($K_{0.5}Na_{0.5}Nb_{0.99}Mn_{0.01}O_3$, denoted as Mn-KNN) was prepared through a modified alkali-based synthesis route reported by Kovacova et al (route A in Ref. [15]). Note that the exact Mn-doping was determined by process optimization of interdependent parameters, i.e., excess of alkali ions, pyrolysis temperature and crystallization temperature. The starting precursors were potassium acetate ($CH_3COOK$, anhydrous, ≥99%, Sigma-Aldrich), sodium acetate ($CH_3COONa$, anhydrous, ≥99%, Sigma-Aldrich), manganese (II) acetate tetrahydrate ($(CH_3COO)_2Mn·4H_2O$, 99.9%, Sigma-Aldrich), and niobium(V) ethoxide ($(CH_3CH_2O)_5Nb$, 99.99%, Thermo Fisher). 2-methoxyethanol (2-MOE, anhydrous, ≥99.8%, Sigma-Aldrich) and acetic acid ($C_2H_4O_2$, glacial, ≥99%, Sigma-Aldrich) were used as solvents, while acetylacetone was used as a modifier. A 10 % excess of potassium acetate and 5 % excess of sodium acetate were added to compensate for the loss of alkalis during thermal annealing. Because these losses are difficult to quantify, we did not include charge compensation for the different valences of $Mn^{2+}$ and $Nb^{5+}$ in the initial formula.

Before synthesis, water was removed from manganese acetate tetrahydrate by freeze-drying. Approximately 20 g of manganese acetate tetrahydrate was placed in a round-bottom flask, which was then immersed in liquid nitrogen for about 30 minutes and subsequently connected to a freeze-drying system (Alpha 3-4 LSCbasic, Christ). Sublimation was carried out at a condenser temperature of −105 °C and a pressure of ~0.93 mbar for 3 days. After completion, the dried manganese acetate was immediately transferred to an Ar-filled glovebox, where weighing and handling of the solutions were performed. 20 mL of 0.1 M manganese acetate stock solution were prepared by dissolving it in acetic acid. All the acetate precursors were weighed and dissolved separately in acetic acid, followed by mixing them together. A clear alkali solution was then transferred to a modified Schlenk apparatus to perform reflux under Ar atmosphere for 10 min at 135 °C, followed by cooling to room temperature. Simultaneously, niobium(V) ethoxide and acetylacetone (2:1 molar ratio to the niobium ion) were dissolved in 2-MOE and were stirred for 30 min at room temperature. This solution was then mixed with the cooled alkali-solution, followed by an additional reflux at 135 °C for 4 h under Ar, and distillation to remove ~50 % of the solvent mixture. After cooling to room temperature, the final solution concentration was adjusted to 0.4 M with the addition of 2-MOE.

It is important to highlight the use of 2-MOE at this stage. Although this solvent is a known carcinogenic and teratogenic agent, it is highly effective in chelation, resulting in stable precursor solutions with a long shelf life. Our recent work on solution-processed PZT films has shown that 1-methoxy-2-propanol can serve as a suitable alternative [20]. This solvent has an almost identical molecular backbone, differing only by a methyl substituent in the α-position of the alcohol moiety. As such, it provides similarly strong chelating capabilities while being considerably less harmful. Another possibility is the use of diol-based solvents.[21] However, exploring the applicability of these approaches for KNN solutions falls outside the scope of this paper.

*2.2 Deposition of the Films*

Mn-KNN thin films were deposited on platinized silicon (Pt/Si, SINTEF) substrates as illustrated in Figure S1a. The substrates were first dehydrated on a hotplate at 400 °C for 1 min. The Mn-KNN solution was spin coated onto the substrates at 3000 rpm for 30 s and then pyrolyzed at 450 °C for 1 min. Spin coating and pyrolysis steps were repeated four times. Crystallization was achieved in a rapid thermal annealing (RTA) furnace (AS-Master,

Annealsys) at 750 °C for 5 min in air with a heating rate of 50 °C/s. The whole process (i.e., four times deposition-pyrolysis and then crystallization) was repeated five times for a total of 20 layers to achieve 1 μm-thick Mn-KNN films on Pt/Si as shown in Figure S1.

### *2.3 Characterization*

The synthesized solution was analysed with inductively coupled plasma optical emission spectroscopy (ICP-OES). The samples were taken using micropipettes, then injected onto a previously calibrated tool (ICAP PRO XP model, Thermofisher). The raw data was evaluated, taking into account the dilutions undertaken, making it possible to return to a concentration in g/L.

The crystalline phase and orientation of Mn-KNN films were investigated with X-ray diffraction (XRD, D8 Discover, Bruker) using Cu-$K_\alpha$ radiation. A locked-couple mode ($\theta$-2$\theta$ scans) was employed in the 2$\theta$ range of 18° - 50° with a step size of 0.02°. Surface and cross-sectional images of the films were obtained with scanning electron microscopy (SEM, Helios NanoLab 650, FEI). 1 nm of Pt was sputtered to decrease charging during SEM analysis (EM ACE600, Leica). Grain size measurements were performed using ImageJ software.

The transmission electron microscopy (TEM) measurements were performed on a JEOL JEM-F200 cold FEG microscope operating at an acceleration voltage of 200 kV. A TEM lamella was prepared following the "lift-out" method with a FEI Helios Nanolab 650 Focused Ion Beam Scanning Electron Microscope (FIB-SEM). Energy Dispersive X-ray Spectroscopy (EDXS), using dual JEOL 100 mm² Silicon Drift Detectors, was performed in scanning mode (STEM) allowing elemental mapping and profiling. Crystalline phase identification was obtained by performing Selected Area Electron Diffraction (SAED) with an aperture allowing to select a 600 nm diameter circular region of interest of the specimen.

100 nm-thick Pt top electrodes were pattern by standard photolithography and a lift-off process. A LOR 3A/S1813 resist bilayer was spin coated on top of the film. Circular shapes with a diameter of 200 µm were exposed with a 405 nm laser wavelength (MLA 150, Heidelberg Instruments). After development, a thin film of 100 nm Pt was sputtered (MED 020 Metallizer, BALTEC), followed by the lift-off. The platinum layer on the platinized silicon substrate served as the bottom electrode.

A post-deposition annealing treatment was performed on a hot plate at 450 °C for 5 min. Polarization $P$ – electric field $E$ loops were measured using a TF Analyzer 2000E (aixACCT) by applying an AC electric field with a triangular waveform at 1 kHz. The electric field dependent dielectric permittivity $\varepsilon_r$, and dielectric losses tan$\delta$ of the films were obtained by applying a DC voltage up to 100 V (staircase mode) coupled with a small-signal AC voltage of 100 mV at 1 kHz. For time-stability tests, the samples were stored in the laboratory at ambient conditions, with a relative humidity ranging from 35 to 50 %.

Converse measurements of the piezoelectric coefficient $e_{31,f}$ were performed by making 3×25×0.675 mm$^3$ cantilevers with Pt top electrodes of 2×10 mm$^2$. The cantilevers were patterned using photolithography and were cut from the wafers with a wire saw. Two types of measurements were done: the large-signal coefficient was measured by employing an AC voltage of ±20 V with a triangular waveform at 13 Hz; the small-signal coefficient was measured by applying a DC-voltage up to 15 V, with a simultaneous small-signal AC voltage of 0.5 V at 100 Hz. Displacement of the cantilever was measured 10 mm from its clamping point. Poisson ratio and Young's modulus of 0.064 and 169 GPa (Si <110> direction), respectively, were used for calculation of mechanical stress and piezoelectric coefficients. The transverse converse piezoelectric coefficient $e_{31,f}$ of the piezoelectric film was calculated as [22]:

$$e_{31,f} = -\frac{1}{3}\frac{Y_{sub}}{(1-\nu_{sub})c_f}\left(c_f + \left(1-c_f\right)(1-\nu_{Sub})\right)\frac{t_{sub}^2}{x_1^2}\left(\frac{w(x_1)}{V}\right), \qquad (1)$$

where $Y_{\text{sub}}$, $\nu_{\text{sub}}$ and $t_{\text{sub}}$ are Young's modulus (169 GPa), Poisson ratio (0.064) and thickness (0.68 mm) of the substrate. $x_1$, $w(x_1)$ and $V$ are measurement position from the clamping (and length of the top electrode, 10 mm), displacement measured at $x_1$ position and applied voltage, respectively. $c_f$ is an electrode coverage factor, which is in this case 0.67 (ratio between the top electrode and cantilever widths). Note that the estimated first resonant frequency of such clamped cantilevers is 3 kHz, which is well above the measurement frequencies.

### *2.4 Haptic Device*

The design of our haptic device was performed by 2D finite element modelling (FEM) (*x*-*z* axes) using COMSOL Multiphysics software. The final device is 680 µm thick, 40 mm long, and 6 mm wide. This optimized design was obtained by considering the resonance of the antisymmetric $A_0$ standing Lamb wave, a suitable size for interaction with a human finger, and low power consumption.[2,23,24] Electrodes were omitted in modelling due to their negligible effect on the overall acoustic properties. The Lamb wave was identified at 19.5 kHz. The device was obtained by depositing 100 nm thick top Pt electrodes on KNN films/Pt/Si. A diamond saw was used to finalize the device. Note that the exact position of the top electrodes is obtained by modelling and is the main parameter to ensure optimized Lamb waves.

The fabricated haptic device was then positioned on two suspended adhesive Kapton strips mounted on a rectangular acrylic frame on a computer-controlled X-Y stage (see our previous publication [2]). A unipolar voltage was applied to the device with a function generator (33210A, Keysight) coupled with a high voltage amplifier (WMA-300, Falco systems). The out-of-plane displacement was measured with a vibrometer (OFV-5000, Polytec). The unipolar input voltage amplitude from the function generator and the output voltage from the haptic device were measured using a digital oscilloscope (TDS2024C, Tektronix). 2D displacement maps of the entire device were acquired through the programming of a the X-Y stage (Labview, National Instruments). A frequency sweep (18 to 22 kHz) of out-of-plane displacements at a specific location on the haptic device was conducted by using a unipolar voltage of 29 V (14.5 $V_{AC}$ + 14.5 $V_{DC}$).

**3. Results and Discussion**

Upon initial processing attempts, large variations in the properties of Mn-KNN films prepared from different solutions were observed, and special care was taken to ensure the exact stoichiometry of the prepared solutions. Especially critical is Mn acetate, which is added in very small amounts: preparing a 50 mL KNN solution with an Mn doping of 1 mol.% requires weighing 0.017 g of freeze-dried Mn (II) acetate, which was impractical. Instead, we prepared first 20 mL of 0.1 M stock solution of Mn acetate in 2-methoxyethanol and added 2 mL of it to the KNN solution using a micropipette. The solution thus prepared was analysed using ICP-OES, and the results are shown in Table 1. The experimentally obtained molar concentrations of individual metals match well the theoretical values, with a slight excess of A-site cations.

Table 1: Results of ICP-OES analysis of the Mn-KNN solution. Atomic concentrations were calculated relative to the perovskite $ABO_3$ structure. Theoretical atomic concentrations correspond to those used for calculation of masses of the precursors during weighing.

| Element | K | Na | Nb | Mn |
|---|---|---|---|---|
| Mass concentration (g/L) | 7.73 | 4.35 | 32.1 | 0.21 |
| Atomic concentration (%) | 11.3 | 10.8 | 19.8 | 0.22 |
| Theoretical atomic concentration (%) | 11.0 | 10.5 | 19.8 | 0.20 |

The XRD pattern of a 20-layer film prepared from the optimized solutions is shown in Figure 1a. It reveals the presence of a perovskite phase with almost random orientation. Notably, peaks at 22.4°, 31.8°, and 45.8°/46.3° correspond to (100), (110), and (002)/(200) planes, respectively. This is in agreement with an orthorhombic (or monoclinic) crystal structure.[13–15,25] All other peaks are coming from the Pt/Si substrate, whose pattern is also shown. Importantly, no peaks corresponding to secondary phases are observed. In Figure 1b, an SEM surface micrograph reveals a dense polycrystalline microstructure. The average grain size, determined by analysing 50 grains and fitting with a normal distribution function, is 52 nm with a standard deviation of 17 nm (see Figure S2). The cross-sectional micrograph (inset of Figure 1b) shows a granular microstructure with an approximate thickness of 1 μm.

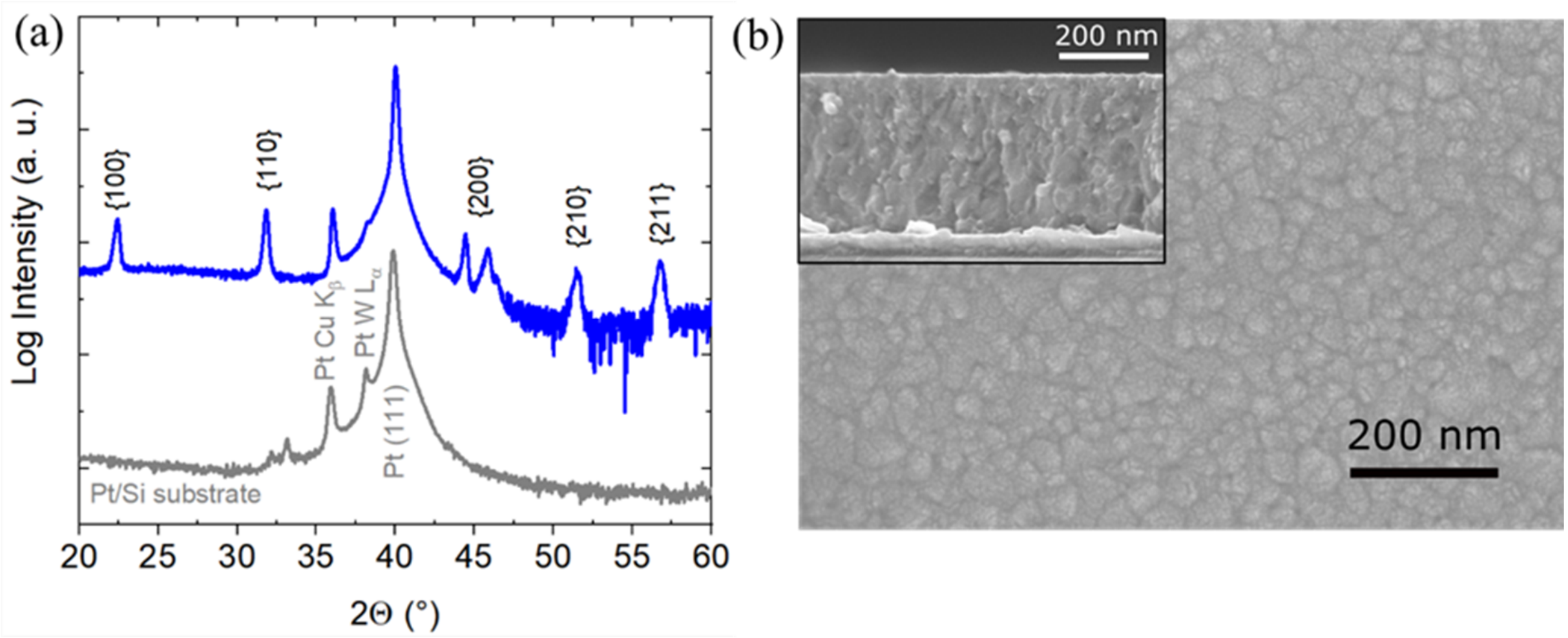

Figure 1: Structural and morphological analysis of Mn-KNN films (20 layers) on a Pt/Si substrate. a) X-ray diffraction pattern. b) SEM surface and cross-sectional (inset) micrographs. In a) peaks of the perovskite phase are denoted with Miller indices. The main peak of the substrate at 39.9 ° corresponds to the Pt (111) peak, while peaks at 36.0 ° and 39.1 ° correspond to Pt Cu $K_\beta$ and Pt W $L_\alpha$, respectively. The peaks in the XRD patterns have been indexed according to ICDD card No. 00-065-0278 for perovskite-structured KNN and ICDD card No. 00-004-0802 for Pt.

The microstructure of the sample was further analysed using TEM. The bright field (BF) micrograph (Figure 2a) confirms ~1 µm thickness of the sample, equiaxed microstructure and grains with diameters ranging from ~15 to ~150 nm. A strong contrast difference between the grains is arising from their different orientation. The film is 100 % dense (no pores), and individual crystallization layers are not distinguishable. A selected area electron diffraction (SAED) pattern, taken across the film thickness, is shown in Figure 2b. The positions of the bright concentrically arranged spots match well the reciprocal-space distances calculated from the ICDD pattern of the orthorhombic phase (Figure 2c). The spotty nature of the pattern is a consequence of large grains (relative to the SAED aperture size) present in the film. However, diffuse broad concentric rings can be observed in the background. The most prominent one has maximum at ~3 $nm^{-1}$ (Figure 2c) and could indicate presence of secondary phases in small quantity.[26,27] The composition of the films was investigated by performing compositional electron-dispersive X-ray spectroscopy (EDXS) line scans across the thickness of the films; the results are shown in Figure 2d. In addition to the elements present in the KNN film, the platinum L line is added for easier determination of top and bottom interfaces. Oxygen, sodium, potassium and niobium K lines all show similar behaviour, i.e., a slight increase of intensity at the surface of the sample, which stabilizes in the bulk of the sample. The observed shape of the line could be due to the non-uniform thickness of the sample. To verify this, we plotted ratios of intensities for different elements (Figure 2e and f). In all cases, straight horizontal lines are obtained across the thickness of the film, confirming the excellent compositional homogeneity of the sample. Note that manganese was analysed also. However, due to its low quantity the obtained signal was too low to draw reliable conclusions. Diffusion of elements between the film and the substrate was not detected. Note that this result is in sharp contrast compared to solution-processed PZT thin films, where compositional gradients are observed unless special process design is performed.[28,29]

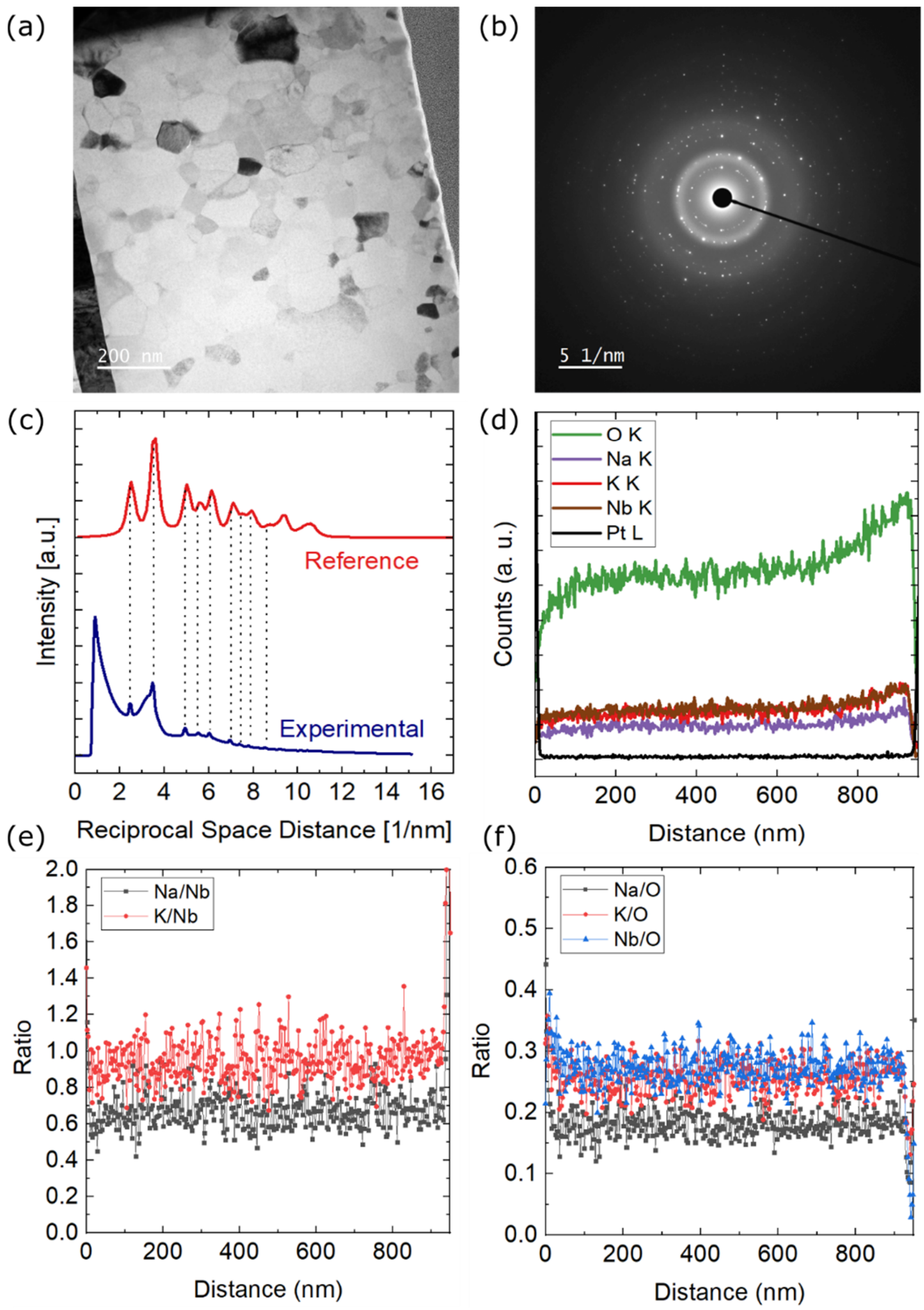

Figure 2: TEM analysis of Mn-KNN film. a) Bright-field (BF) cross-sectional micrograph. b) Selected area electron diffraction (SAED) taken over the central part of the micrograph shown in a). c) Line scan intensity profile measured on SAED pattern (azimuthal average, performed with DiffTools script [30]) compared with the theoretical pattern of the orthorhombic perovskite phase (Ref. ICDD pattern No. 00-065-0278 [31]). d) Energy-dispersive X-ray spectroscopy (EDXS) line scans across thickness of the sample. The Pt L signal is added to determine the top and bottom interface of the film. e) Ratios of Na K and K

K signals against Nb K signal. (f) Ratios of Na K, K K, and Nb K against O K signal. Zero distance signifies interface of the film with the bottom electrode.

The room temperature polarization $P$ - electric field $E$ loops are shown in Figure 3a. Slim $P$-$E$ loops are obtained, and the resulting remanent polarization $P_r$ and coercive field $E_c$ values are 10 μC/cm$^2$ and 40 kV/cm, respectively. The values are comparable to state-of-the-art values reported for doped KNN films produced via CSD (Figure 3b). Matsuda et al.[16] reported slim $P$-$E$ loops for a 1 μm thick film with 1 mol.% of Mn-KNN, exhibiting a $P_r$ value of 14 μC/cm$^2$, and a higher $E_c$ value of 97.5 kV/cm. Wang et.al.[17] reported a 3.5 μm-thick non-doped KNN film using a polyvinylpyrrolidone-modified CSD process, with a higher $P_r$ of 16.4 μC/cm$^2$ and a lower $E_c$ of 42 kV/cm. However, these loops were not saturated, and the breakdown field was ±600 kV/cm. When they doped the films with 2 % of Mn, a moderate $P_r$ of 10 μC/cm$^2$ and $E_c$ of 40 kV/cm were achieved. Doped films thus sustained higher electric fields up to at least 1 MV/cm.

Transient current $j$ loops (Figure 3a) of our Mn-KNN films show that leakage current contribution remains low even at a field as high as 1 MV/cm (see Figure S3 for detailed electric-field dependence of both $P$ and $j$). Low breakdown fields and increased leakage current at higher applied electric fields are often reported in KNN films and are typically attributed to a high concentration of oxygen vacancies and hole charge carriers, both being caused by the loss of alkali-ions upon annealing.[7,11–18,32] We attribute the excellent dielectric behaviour of our films to efficient charge compensation achieved through doping with Mn. Note also that the ferroelectric response remains stable for at least 7 months (Figure 3a), which is further confirming the excellent quality of the films.

The dependency of permittivity $\varepsilon_r$ of the Mn-KNN film on an electrical dc bias field shows the typical response for ferroelectric films, as illustrated in Figure 3c. The permittivity value decreases from 920 to 70 as the electric field increases from 0 to 1 MV/cm, which leads to a tunability $\eta = (\varepsilon(0) - \varepsilon(E))/(\varepsilon(0))$ of 0.92. This is comparable to (001)-oriented 500 nm-thick epitaxial KNN films, in which a tunability of 0.81 was achieved at 780 kV/cm.[33] The dielectric losses tan$\delta$ remain below 0.06 at all measured fields. Reproducibility is a common issue in KNN material.[34,35] To assess it, properties of two films prepared from two different solutions were measured (Figure S4). Both films exhibit very similar ferroelectric characteristics, confirming good reproducibility of the process.

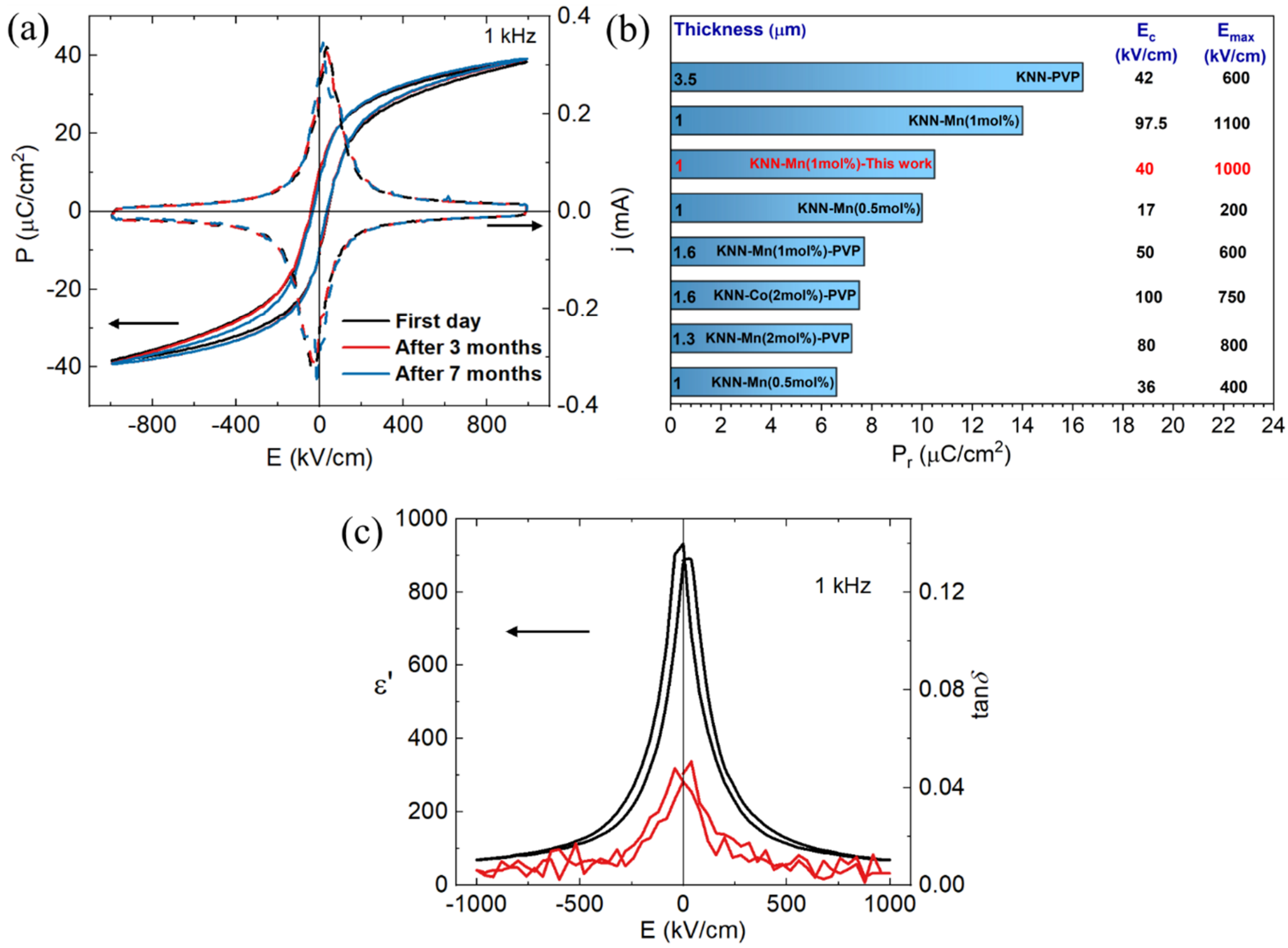

Figure 3: Ferroelectric characterization of 1 μm-thick Mn-KNN film. a) Polarization $P$ and transient current $j$ as functions of electric field $E$. b) Comparison of remanent polarization $P_r$, coercive field $E_c$, and maximum electric field ($E_{max}$) of the film reported in this work and other

CSD-based KNN films from the literature with thicknesses between 1 and 3.5 μm.[12,14–19] c) Electric field *E* dependent relative permittivity $\varepsilon_r$ and dielectric loss tan$\delta$.

To assess piezoelectric properties, the transverse piezoelectric coefficient $e_{31,f}$ was measured. In Figure 4a, the displacement of a cantilever with 1 μm Mn-KNN is shown as a function of the applied electric field. At 200 kV/cm the displacement reaches 500 nm, which results in an estimated $e_{31,f}$ large-signal coefficient of -15.4 C/m$^2$. Note that a slight shift of the loop towards negative field values might result from the internal field generated by the charged defect dipoles within the film, suggesting that further optimization of charge compensation in the film is still required. The small-signal $e_{31,f}$ coefficient was also measured (Figure 4b), and its maximum value is -12.6 C/m$^2$, closely aligning with the large signal value. As shown in Figure 4c, the $e_{31,f}$ value of the Mn-KNN film surpasses those reported for hydrothermally grown, pulsed laser deposited (PLD), and CSD-derived KNN films and is comparable to that of Shibata's sputtered KNN films.[11,12,36,37] Such high response can be attributed to dense microstructure and compositional homogeneity of the films.

The dielectric, ferroelectric and piezoelectric properties of the Mn-KNN films are collected and compared to state-of-the-art PZT films in Table S1. While Mn-KNN films show lower remanent polarization values, i.e., 10 vs. 38 μC/cm$^2$, the piezoelectric coefficient $e_{31,f}$ falls into the same range as for PZT films. Note also that the $e_{31,f}$ value is comparable to the highest values reported for sputtered lead-free $0.93Na_{0.5}Bi_{0.5}TiO_3$-$0.07BaTiO_3$ (NBT-BT) films.[38]

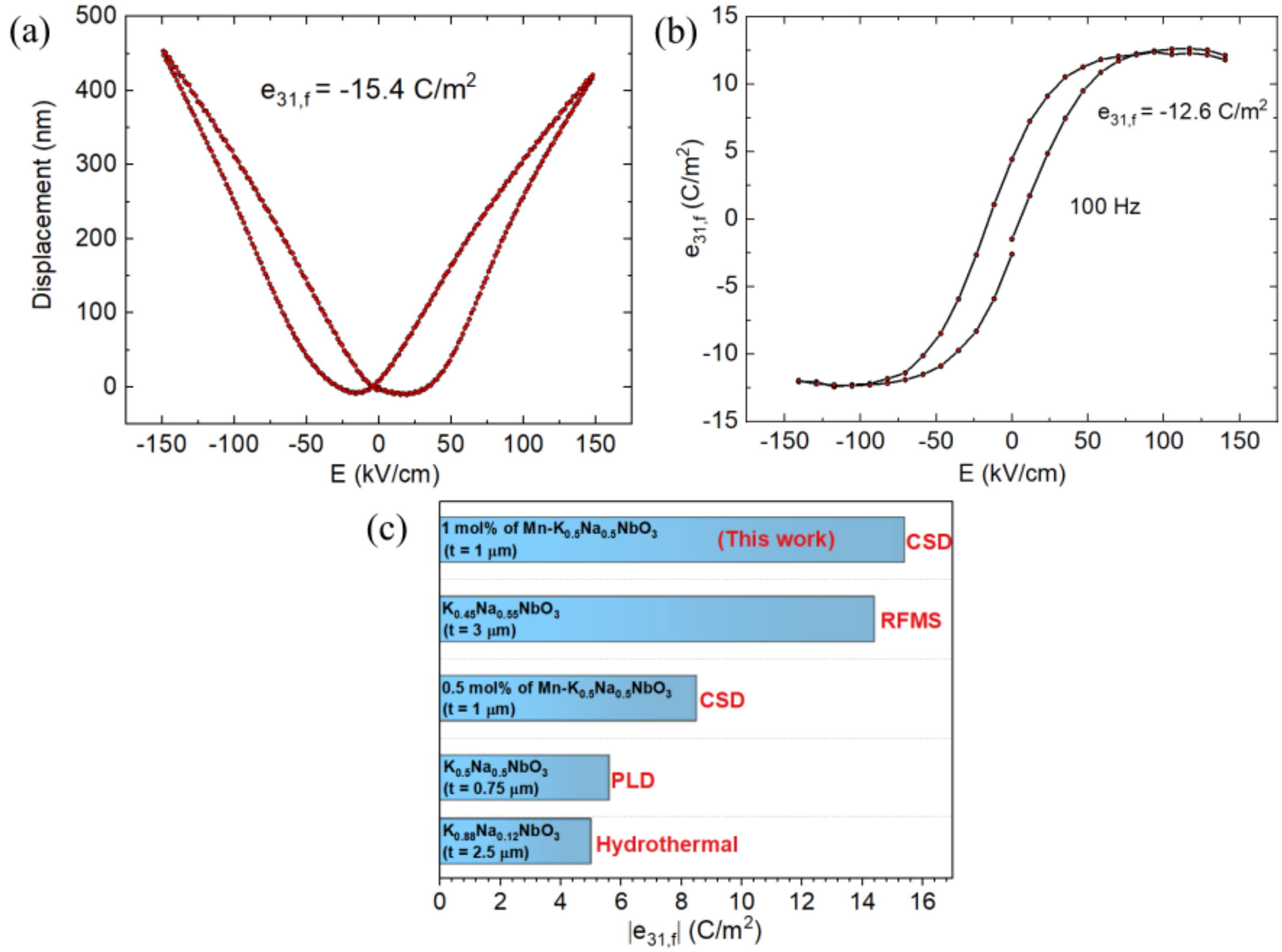

Figure 4: Transverse piezoelectric coefficient $e_{31,f}$ of 1 μm-thick Mn-KNN film using the cantilever method. a) Large-signal displacement obtained at 18 Hz. b) Small-signal $e_{31,f}$ loop obtained at 100 Hz. c) $e_{31,f}$ value of the Mn-KNN film compared to values reported for KNN prepared by hydrothermal method, pulsed laser deposition (PLD), chemical solution deposition (CSD), and radio-frequency magnetron sputtering (RFMS). [11,12,36,37]

A piezoelectric haptic device was fabricated on a Pt/Si substrate, and its design is schematically shown in Figure 5a. The device is 40 mm long and 6 mm wide, and two 1 x 4 mm² metal-insulator-metal (MIM) actuators were integrated to generate acoustic waves within the structure. The capacitance of each actuator was ~13 nF (see Figure S6).

To actuate the device, the two actuators were connected in series and driven with a unipolar sinusoidal voltage. The vibrometer laser was positioned at one of the anti-nodes, where the out-of-plane displacement reaches its maximum. The frequency-dependent response is presented in Figure 5b; a unipolar voltage of 29 V (comprising 14.5 $V_{AC}$ + 14.5 $V_{DC}$) was applied. The resonant frequency ($f_r$) was observed at 19.5 kHz, as predicted with FEM simulation. The quality factor, defined as $f_r/\Delta f$, where $\Delta f$ is the full width at half maximum (FWHM) of the resonance peak, was calculated to be 17.2.

The displacement was then scanned along the length of the device, revealing an acoustic wave with three anti-nodes, which agrees with the FEM simulation shown in Figure 5c. The displacement was further measured as a function of the applied unipolar voltage (Figure 5d). It increases with rising voltage and becomes approximately linear above ~15 V (150 kV/cm). The displacement required for practical operation, i.e., 1 μm, was achieved at 21 V. To further confirm the nature of the acoustic wave, a two-dimensional displacement map was acquired across the entire device surface and is presented in Figure 5e.

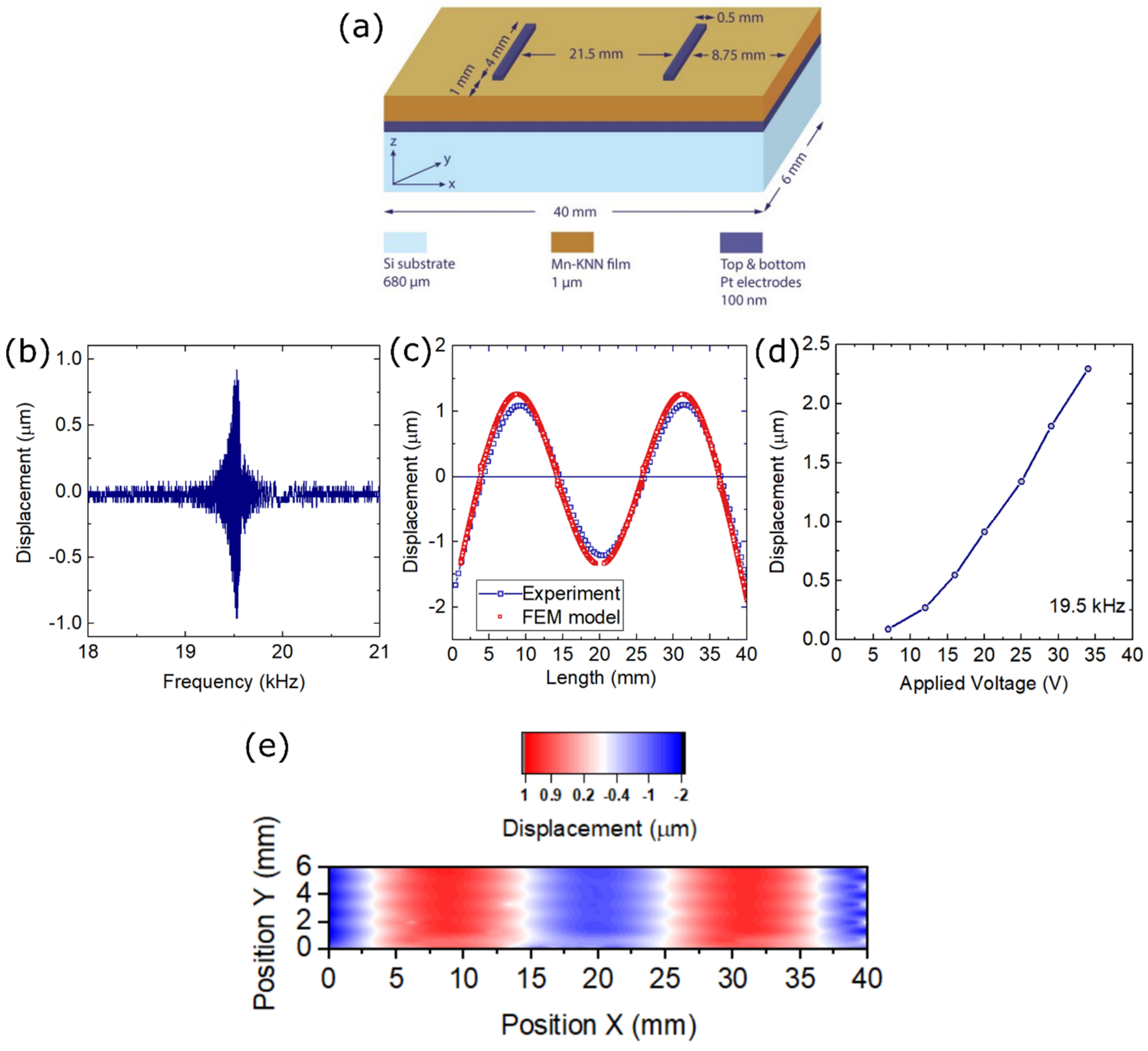

Figure 5: Lead-free piezoelectric surface haptic device. a) Schematic of the device structure. b) Out-of-plane displacement as a function of frequency, measured at the anti-node under a unipolar excitation voltage of 29 V (14.5 $V_{AC}$ + 14.5 $V_{DC}$). c) Experimental and simulated displacement profiles along the length of the device. d) Unipolar voltage dependence of the out-of-plane displacement measured at the anti-node. e) Two-dimensional displacement map across the entire device surface at 19.5 kHz under a unipolar voltage of 29 V.

From a device perspective, it is important to estimate the power consumption, $P$, which can be calculated using the applied RMS voltage $V_{rms}$, the measured RMS current ($I_{rms}$) through the actuators, and the phase shift ($\theta$) between them, according to the relation: $P = V_{rms} \cdot I_{rms} \cdot \cos\theta$. At the resonant frequency of 19.5 kHz, and under an applied unipolar voltage of 25 V ($V_{rms}$ = 15.3 V), the measured power consumption was 75 mW. This value is either lower than or comparable to the power consumption reported for other piezoelectric surface haptic devices based on PZT and AlN films in a metal-insulator-metal (MIM) configuration (see Table 2). Compared to the power consumption of 1216 mW recently reported for a transparent KNN-based haptic device on glass, the power consumption reported here is significantly lower. This difference can be attributed to the superior KNN material quality on Pt/Si, as well as differences in design, including geometry and substrate material [39].

Table 2: Performance of the film-based metal-insulator-metal (MIM) piezoelectric haptic devices reported in the literature. $t_f$ – thickness of the piezoelectric film; N – number of actuators on the device; $f_r$ - resonance frequency; $V_{app}$ – voltage applied to the device; $C_A$ – capacitance of a single actuator; $P$ – power consumption of the device.

| Device stack | Size ($mm^3$) | $t_f$ (μm) | N | $f_r$ (kHz) | $V_{app}$ (V) | $C_A$ (nF) | $P$ (mW) | Ref. |
|---|---|---|---|---|---|---|---|---|
| Mo/AIN/Mo/Glass | 110×65×0.7 | 2 | 30 | 23 | 60 | 1.1 | 3000 | [40] |
| ITO/PZT/ITO/Glass | 60×71×0.5 | 2 | 30 | 22.5 | 10 | 18 | 3580 | [4] |
| Pt/PZT/Pt/Si | 34×10×0.65 | 0.5 | 3 | 71.5 | 40 | 160 | 2300 | [23] |
| Pt/PZT/Pt/Si | 34×10×0.65 | 10 | 3 | 62 | 40 | 12 | 150 | [23] |
| Pt/PZT-Nb/Pt/$Al_2O_3$ | 25.5×12.5×0.25 | 80 | 2 | 17.8 | 20 | 1 | 5 | [24] |
| ITO/KNN-Mn/FTO/glass | 50×10×0.3 | 0.6 | 4 | 16.5 | 112 | 64 | 1216 | [39] |
| Pt/KNN-Mn/Pt/Si | 40×6×0.68 | 1 | 2 | 19.5 | 25 | 13 | 75 | This work |

## Conclusions

1 μm-thick lead-free polycrystalline Mn-doped KNN films (1 mol.% Mn) were fabricated using chemical solution deposition. The films exhibit a dense granular microstructure with a single-phase perovskite crystal structure and random orientation. They show robust ferroelectric performance, including a remanent polarization of 10 μC/cm$^2$, a reduced coercive field of 40 kV/cm, and low leakage current even at electric fields up to 1 MV/cm. These properties are highly reproducible and remain stable for at least seven months.

The films also demonstrate excellent piezoelectric ($e_{31,\mathrm{f}}$ = -15.4 C/m$^2$) and dielectric properties ($\varepsilon_\mathrm{r}$ = 921, tan$\sigma \leq$ 0.06). The measured $e_{31,\mathrm{f}}$ values exceed those reported for other CSD-derived KNN films and are comparable to those of sputtered KNN and PZT thin films. We attribute the enhanced ferroelectric and piezoelectric performance to the precise control of the film composition, specifically the combination of Mn doping together with 10% potassium and 5% sodium excess.

Finally, a surface-haptic device incorporating Mn-KNN film actuators delivers a clear response above 1 μm displacement at 25 V with a power consumption of only 75 mW, demonstrating the strong potential of Mn-KNN films for piezoelectric applications.

**Acknowledgements**

B. Mandal and S. Glinsek acknowledge Luxembourg National Research Fund (FNR) for financial support through the project INTER/NWO/20/15079143/TRICOLOR. J. Cardoletti and S. Glinsek acknowledge FNR for financial support through the project FLASHPOX (C21/MS/16215707). FILAB laboratory is acknowledged for performing ICP-OES analysis.

**Conflict of Interest**

Authors declare that there are no conflicts of interest.

**Declaration of Generative AI and AI-assisted technologies in the writing process**

During the preparation of this work the authors used ChatGPT to improve English language. After using this tool, the authors reviewed and edited the content as needed and take full responsibility for the content of the publication.

**Data Availability Statement**

The raw data generated in this study have been deposited in the Zenodo repository under DOI: 10.5281/zenodo.17984369.

## References

[1] I. Kanno, J. Ouyang, J. Akedo, T. Yoshimura, B. Malič, P. Muralt, Piezoelectric thin films for MEMS, Appl. Phys. Lett. 122 (2023) 090401. https://doi.org/10.1063/5.0146681.

[2] S. Glinsek, M.A. Mahjoub, M. Rupin, T. Schenk, N. Godard, S. Girod, J.-B. Chemin, R. Leturcq, N. Valle, S. Klein, C. Chappaz, E. Defay, Fully transparent friction-modulation haptic device based on piezoelectric thin film, Adv. Funct. Mater. 30 (2020) 2003539. https://doi.org/10.1002/adfm.202003539.

[3] P. Muralt, Recent progress in materials issues for piezoelectric MEMS, J. Am. Ceram. Soc. 91 (2008) 1385–1396. https://doi.org/10.1111/j.1551-2916.2008.02421.x.

[4] H. Hua, Y. Chen, Y. Tao, D. Qi, Y. Li, A highly transparent haptic device with an extremely low driving voltage based on piezoelectric PZT films on glass, Sensors Actuators A Phys. 335 (2022) 113396. https://doi.org/10.1016/j.sna.2022.113396.

[5] F. Pavageau, C. Dieppedale, P. Perreau, R. Liechti, A. Hamelin, C. Licitra, F. Casset, G. Le Rhun, Highly transparent piezoelectric PZT membranes for transducer applications, Sensors Actuators A Phys. 346 (2022) 113866. https://doi.org/10.1016/J.SNA.2022.113866.

[6] European Commission, DIRECTIVE 2011/65/EU OF THE EUROPEAN PARLIAMENT AND OF THE COUNCIL of 8 June 2011 - ROHS, Off. J. Eur. Union. 54 (2011) 88–110. https://doi.org/10.1017/CBO9781107415324.004.

[7] J. Wu, D. Xiao, J. Zhu, Potassium–sodium niobate lead-free piezoelectric materials: past, present, and future of phase boundaries, Chem. Rev. 115 (2015) 2559–2595. https://doi.org/10.1021/cr5006809.

[8] T.R. Shrout, S.J. Zhang, Lead-free piezoelectric ceramics: Alternatives for PZT?, J. Electroceramics. 19 (2007) 111–124. https://doi.org/10.1007/s10832-007-9047-0.

[9] W. Liu, X. Ren, Large piezoelectric effect in Pb-free ceramics, Phys. Rev. Lett. 103 (2009) 1–4. https://doi.org/10.1103/PhysRevLett.103.257602.

[10] K. Shibata, R. Wang, T. Tou, J. Koruza, Applications of lead-free piezoelectric materials, MRS Bull. 43 (2018) 612–616. https://doi.org/10.1557/mrs.2018.180.

[11] K. Shibata, K. Suenaga, K. Watanabe, F. Horikiri, A. Nomoto, T. Mishima, Improvement of piezoelectric properties of (K,Na)$NbO_3$ films deposited by sputtering, Jpn. J. Appl. Phys. 50 (2011) 41503. https://doi.org/10.1143/JJAP.50.041503.

[12] S.S. Won, J. Lee, V. Venugopal, D.-J. Kim, J. Lee, I.W. Kim, A.I. Kingon, S.-H. Kim, Lead-free Mn-doped ($K_{0.5}$,$Na_{0.5}$)$NbO_3$ piezoelectric thin films for MEMS-based vibrational energy harvester applications, Appl. Phys. Lett. 108 (2016) 232908. https://doi.org/10.1063/1.4953623.

[13] A. Chowdhury, J. Bould, M.G.S. Londesborough, S.J. Milne, The effect of refluxing on the alkoxide-based sodium potassium niobate solgel system: Thermal and spectroscopic studies, J. Solid State Chem. 184 (2011) 317–324. https://doi.org/10.1016/j.jssc.2010.12.002.

[14] L. Wang, W. Ren, P. Shi, X. Wu, Structures, electrical properties, and leakage current behaviors of un-doped and Mn-doped lead-free ferroelectric $K_{0.5}Na_{0.5}NbO_3$ films, J. Appl. Phys. 115 (2014) 034103. https://doi.org/10.1063/1.4861415.

[15] V. Kovacova, J.I. Yang, L. Jacques, S.W. Ko, W. Zhu, S. Trolier-McKinstry, Comparative solution synthesis of Mn doped (Na,K)$NbO_3$ thin films, Chem. Eur. J. 26 (2020) 9356–9364. https://doi.org/10.1002/chem.202000537.

[16] T. Matsuda, W. Sakamoto, B.-Y. Lee, T. Iijima, J. Kumagai, M. Moriya, T. Yogo, Electrical properties of lead-free ferroelectric Mn-doped $K_{0.5}Na_{0.5}NbO_3$–$CaZrO_3$ thin films prepared by chemical solution deposition, Jpn. J. Appl. Phys. 51 (2012) 09LA03. https://doi.org/10.1143/JJAP.51.09LA03.

[17] L.Y. Wang, K. Yao, W. Ren, Piezoelectric $K_{0.5}Na_{0.5}O_3$ thick films derived from polyvinylpyrrolidone-modified chemical solution deposition, Appl. Phys. Lett. 93 (2008) 92903–92906. https://doi.org/10.1063/1.2978160.

[18] L. Wang, W. Ren, P.C. Goh, K. Yao, P. Shi, X. Wu, X. Yao, Structures and electrical properties of Mn- and Co-doped lead-free ferroelectric $K_{0.5}Na_{0.5}NbO_3$ films prepared by a chemical solution deposition method, Thin Solid Films. 537 (2013) 65–69. https://doi.org/10.1016/j.tsf.2013.04.045.

[19] L. Wang, W. Ren, P. Shi, X. Chen, X. Wu, X. Yao, Enhanced ferroelectric properties in Mn-doped $K_{0.5}Na_{0.5}O_3$ thin films derived from chemical solution deposition, Appl. Phys. Lett. 97 (2010) 072902. https://doi.org/10.1063/1.3479530.

[20] N. Godard, P. Grysan, E. Defay, S. Glinšek, Growth of {100}-oriented lead zirconate titanate thin films mediated by a safe solvent, J. Mater. Chem. C. 9 (2021) 281–287. https://doi.org/10.1039/D0TC04066E.

[21] M.L. Calzada, A. González, R. Jiménez, C. Alemany, J. Mendiola, Rapid thermal processing of strontium bismuth tantalate ferroelectric thin films prepared by a novel chemical solution deposition method, J. Eur. Ceram. Soc. 21 (2001) 1517–1520. https://doi.org/https://doi.org/10.1016/S0955-2219(01)00054-1.

[22] A. Mazzalai, D. Balma, N. Chidambaram, R. Matloub, P. Muralt, Characterization and fatigue of the converse piezoelectric effect in PZT films for MEMS applications, J. Microelectromechanical Syst. 24 (2015) 831–838.

https://doi.org/10.1109/JMEMS.2014.2353855.

[23] L. Song, S. Glinsek, S. Drnovsek, V. Kovacova, B. Malic, E. Defay, Piezoelectric thick film for power-efficient haptic actuator, Appl. Phys. Lett. 121 (2022) 212901. https://doi.org/10.1063/5.0106174/2834598.

[24] L. Song, S. Glinsek, H. Ursic, S. Drnovsek, M. Sadl, B. Malic, E. Defay, Power-efficient piezoelectric haptic actuators with large deflection, Sensors Actuators A Phys. 356 (2023) 114346. https://doi.org/10.1016/j.sna.2023.114346.

[25] J. Tellier, B. Malič, B. Dkhil, D. Jenko, J. Cilenšek, M. Kosec, Crystal structure and phase transitions of sodium potassium niobate perovskites, Solid State Sci. 11 (2009) 320–324. https://doi.org/10.1016/j.solidstatesciences.2008.07.011.

[26] D. Jenko, A. Benčan, B. Malič, J. Holc, M. Kosec, Electron microscopy studies of potassium sodium niobate ceramics, Microsc. Microanal. 11 (2005) 572–580. https://doi.org/10.3390/ma8125449.

[27] F. Madaro, R. Sæterli, J.R. Tolchard, M.-A. Einarsrud, R. Holmestad, T. Grande, Molten salt synthesis of $K_4Nb_6O_{17}$, $K_2Nb_4O_{11}$ and $KNb_3O_8$ crystals with needle-or plate-like morphology, CrystEngComm. 13 (2011) 1304–1313. https://doi.org/10.1039/C0CE00413H.

[28] N. Ledermann, P. Muralt, J. Baborowski, S. Gentil, K. Mukati, M. Cantoni, A. Seifert, N. Setter, {100}-textured, piezoelectric $Pb(Zr_x,Ti_{1-x})O_3$ thin films for MEMS: integration, deposition and properties, Sensors Actuators A Phys. 105 (2003) 162–170. https://doi.org/10.1016/S0924-4247(03)00090-6.

[29] F. Calame, P. Muralt, Growth and properties of gradient free sol-gel lead zirconate titanate thin films, Appl. Phys. Lett. 90 (2007) 62907.

https://doi.org/10.1063/1.2337362.

[30] D.R.G. Mitchell, DiffTools: Electron diffraction software tools for DigitalMicrograph$^{TM}$, Microsc. Res. Tech. 71 (2008) 588–593. https://doi.org/https://doi.org/10.1002/jemt.20591.

[31] ICDD database PDF4+ v4.19, (2019).

[32] L. Jacques, V. Kovacova, J.I. Yang, S. Trolier-McKinstry, Activation energies for crystallization of manganese-doped (K,Na)$NbO_3$ thin films deposited from a chemical solution, J. Am. Ceram. Soc. 104 (2021) 4968–4976. https://doi.org/10.1111/jace.17915.

[33] L. Hao, Y. Yang, Y. Huan, H. Cheng, Y.-Y. Zhao, Y. Wang, J. Yan, W. Ren, J. Ouyang, Achieving a high dielectric tunability in strain-engineered tetragonal $K_{0.5}Na_{0.5}O_3$ films, Npj Comput. Mater. 7 (2021) 62. https://doi.org/10.1038/s41524-021-00528-2.

[34] A. Chowdhury, J. Bould, M.G.S. Londesborough, S.J. Milne, Fundamental issues in the synthesis of ferroelectric $Na_{0.5}K_{0.5}NbO_3$ thin films by sol−gel processing, Chem. Mater. 22 (2010) 3862–3874. https://doi.org/10.1021/cm903697j.

[35] F. Söderlind, P.-O. Käll, U. Helmersson, Sol–gel synthesis and characterization of $Na_{0.5}K_{0.5}NbO_3$ thin films, J. Cryst. Growth. 281 (2005) 468–474. https://doi.org/10.1016/j.jcrysgro.2005.04.044.

[36] M.D. Nguyen, M. Dekkers, E.P. Houwman, H.T. Vu, H.N. Vu, G. Rijnders, Lead-free $K_{0.5}Na_{0.5}O_3$ thin films by pulsed laser deposition driving MEMS-based piezoelectric cantilevers, Mater. Lett. 164 (2016) 413–416. https://doi.org/10.1016/j.matlet.2015.11.044.

[37] A. Tateyama, Y. Ito, T. Shiraishi, Y. Orino, M. Kurosawa, H. Funakubo, Thermal stability of self-polarization in a (K,Na)$NbO_3$ film prepared by the hydrothermal method, Jpn. J. Appl. Phys. 60 (2021) SFFB03. https://doi.org/10.35848/1347-4065/ac10f8.

[38] T. Harigai, Y. Tanaka, H. Adachi, E. Fujii, Piezoelectric properties of lead-free (Na,Bi)$TiO_3$–$BaTiO_3$ (001) epitaxial thin films around the morphotropic phase boundary, Appl. Phys. Express. 3 (2010) 111501. https://doi.org/10.1143/APEX.3.111501.

[39] L. Song, S. Glinsek, N.R. Alluri, V. Kovacova, M. Melchiorr, A.B. Martinez, B. Mandal, J. Cardoletti, E. Defay, Highly transparent lead-free piezoelectric haptic device, Commun. Mater. 6 (2025) 91. https://doi.org/10.1038/s43246-025-00817-z.

[40] F. Casset, J.S. Danel, P. Renaux, C. Chappaz, F. Bernard, T. Sednaoui, S. Basrour, B. Desloges, S. Fanget, 4-inch transparent plates based on thin-film AlN actuators for haptic applications, Mechatronics. 40 (2016) 264–269. https://doi.org/10.1016/j.mechatronics.2016.05.014.

# Supplementary Material for:

# Solution-Based Fabrication of High-performance $K_{0.5}Na_{0.5}NbO_3$ Thin Films for Surface Haptics

Nagamalleswara Rao Alluri,[a,b,*] Longfei Song,[a,b,†] Stephanie Girod,[a] Barnik Mandal,[a,c] Juliette Cardoletti,[a,‡] Vid Bobnar,[d] Torsten Granzow,[a,b] Veronika Kovacova,[a,b] Adrian-Marie Philippe,[e] Emmanuel Defay,[a,b,c] and Sebastjan Glinsek[a,b]

[a]Smart Materials Unit, Luxembourg Institute of Science and Technology, 41 rue du Brill, L-4422 Belvaux, Luxembourg

[b]Inter-Institutional Research Group Uni.lu–LIST on Ferroic Materials, 41 rue du Brill, L-4422 Belvaux, Luxembourg

[c]Department of Physics and Materials Science, University of Luxembourg, 41 rue du Brill, L-4422 Belvaux, Luxembourg

[d]Department of Condensed Matter Physics, Jožef Stefan Institute, Jamova cesta 39, SI-1000 Ljubljana, Slovenia

[e]Advanced Analyses and Support Unit, Luxembourg Institute of Science and Technology, 41 rue du Brill, L-4422 Belvaux, Luxembourg

[*] Currently at Materials Science and Metallurgical Engineering, Kyungpook National University, 80, Daehak-ro, Buk-gu, Daegu, 41566, South Korea

[†] Currently at College of Intelligent Systems, Science and Engineering, Harbin Engineering University, Harbin, 150001, China

[‡] Currently at Institute of Materials Science, Technische Universität Darmstadt, Peter-Grünberg-Strase 2, Darmstadt, 64287, Germany

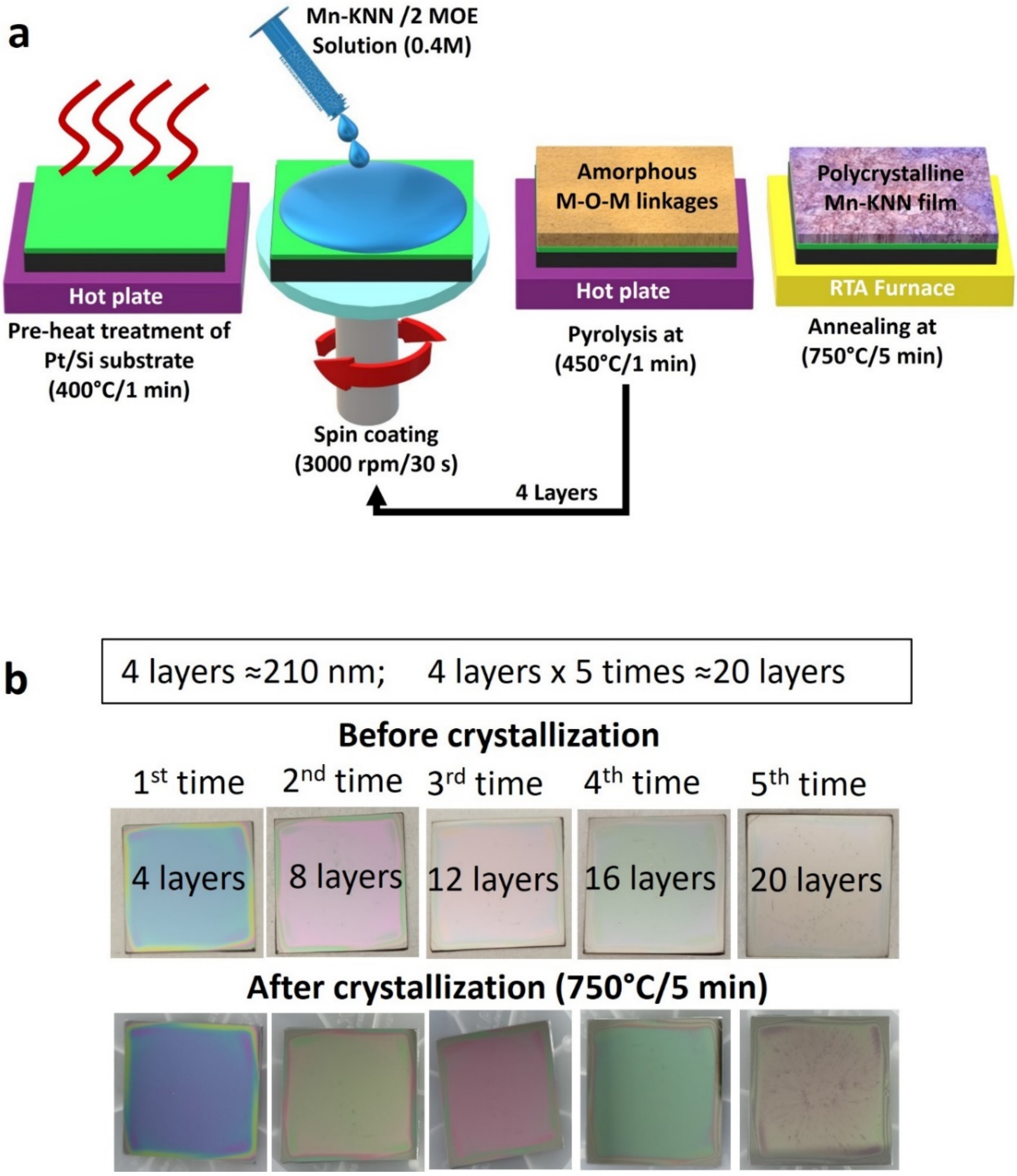

Figure S1: **(a)** Processing steps for fabrication of 1 mol.% of Mn-doped $K_{0.5}Na_{0.5}NbO_3$ (Mn-KNN) films. **(b)** Optical images of the films (4 layers × 5 = 20 layers) before and after crystallization.

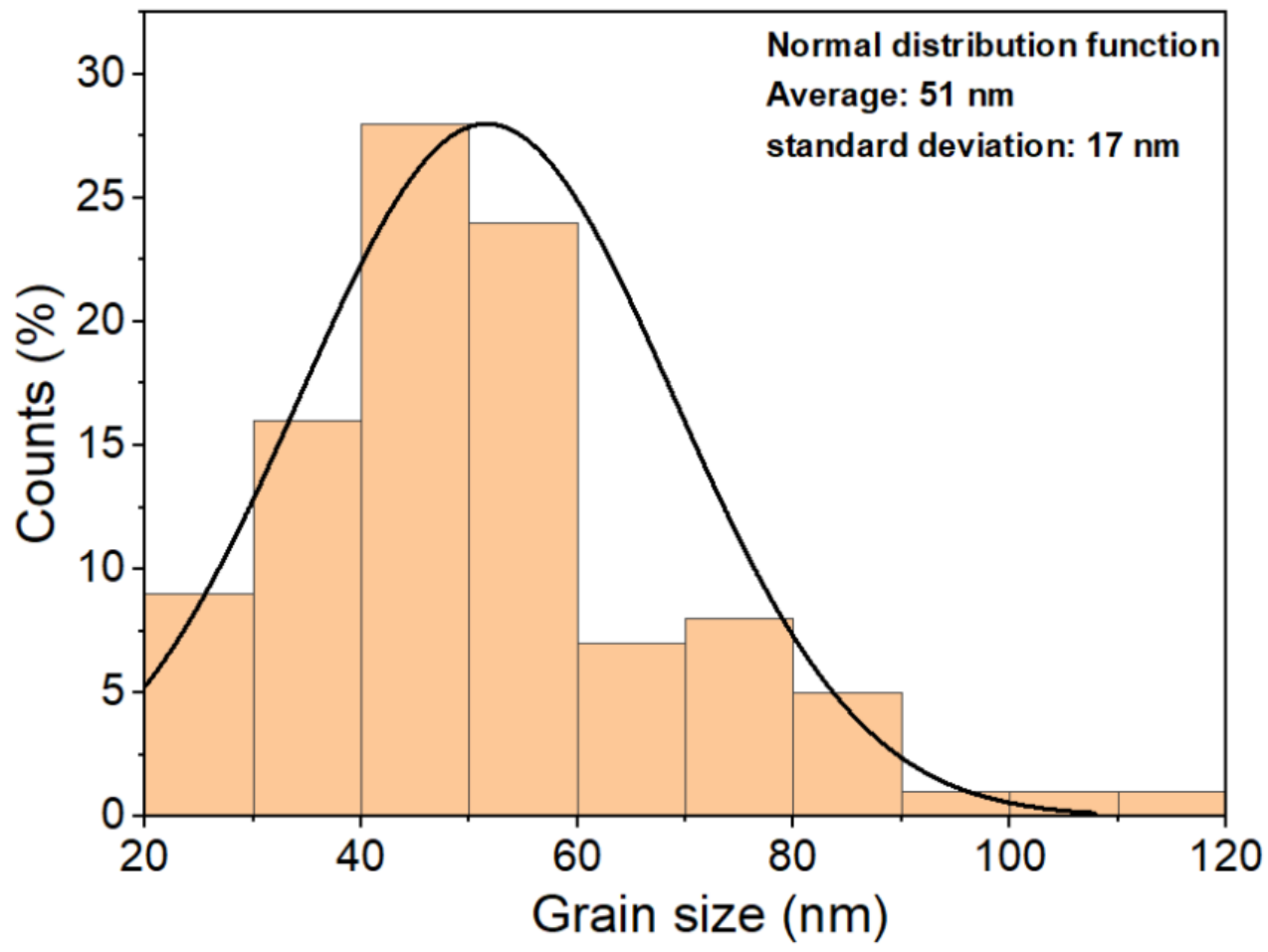

Figure S2: Evaluation of average grain size of Mn-KNN film (20 layers) with ImageJ software by counting 50 grains in SEM micrograph.

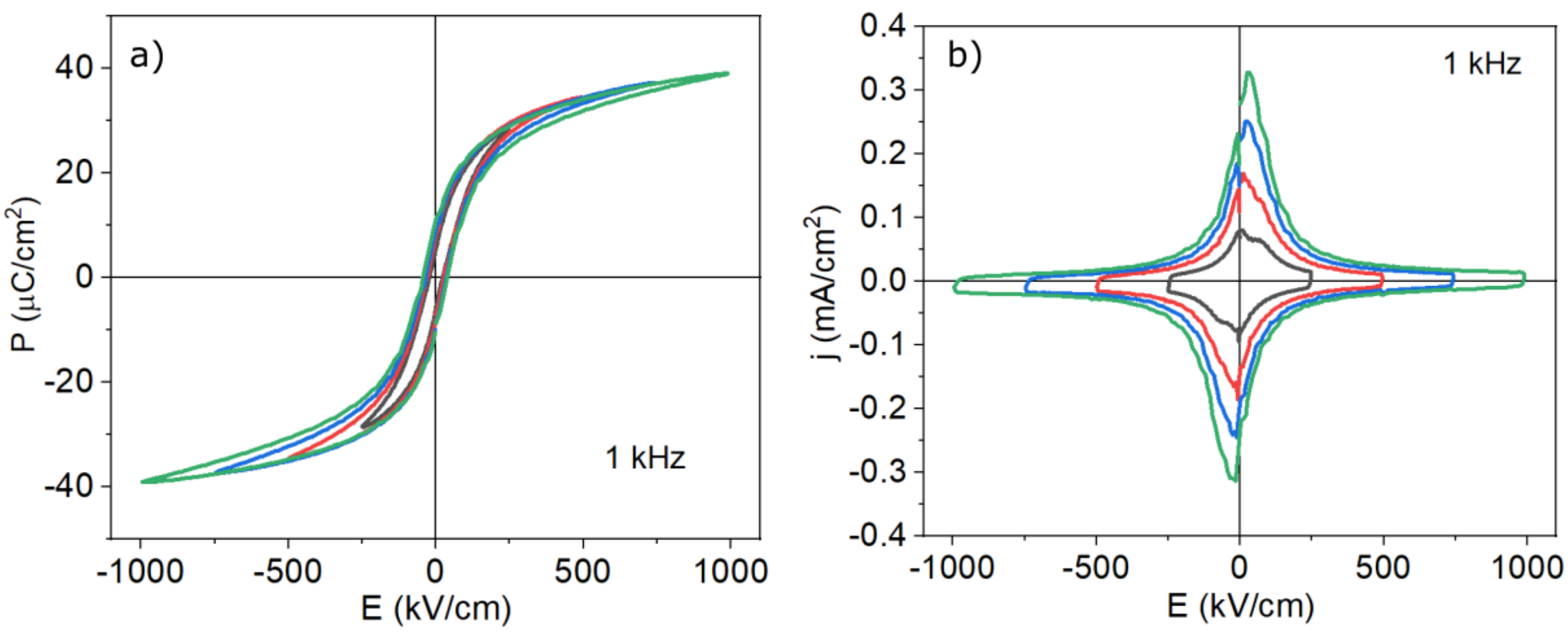

Figure S3: **(a)** Electric-field *E* dependent polarization *P* loops of 1 µm-thick Mn-KNN film. **(b)** Electric-field *E* dependent transient current *j* loops.

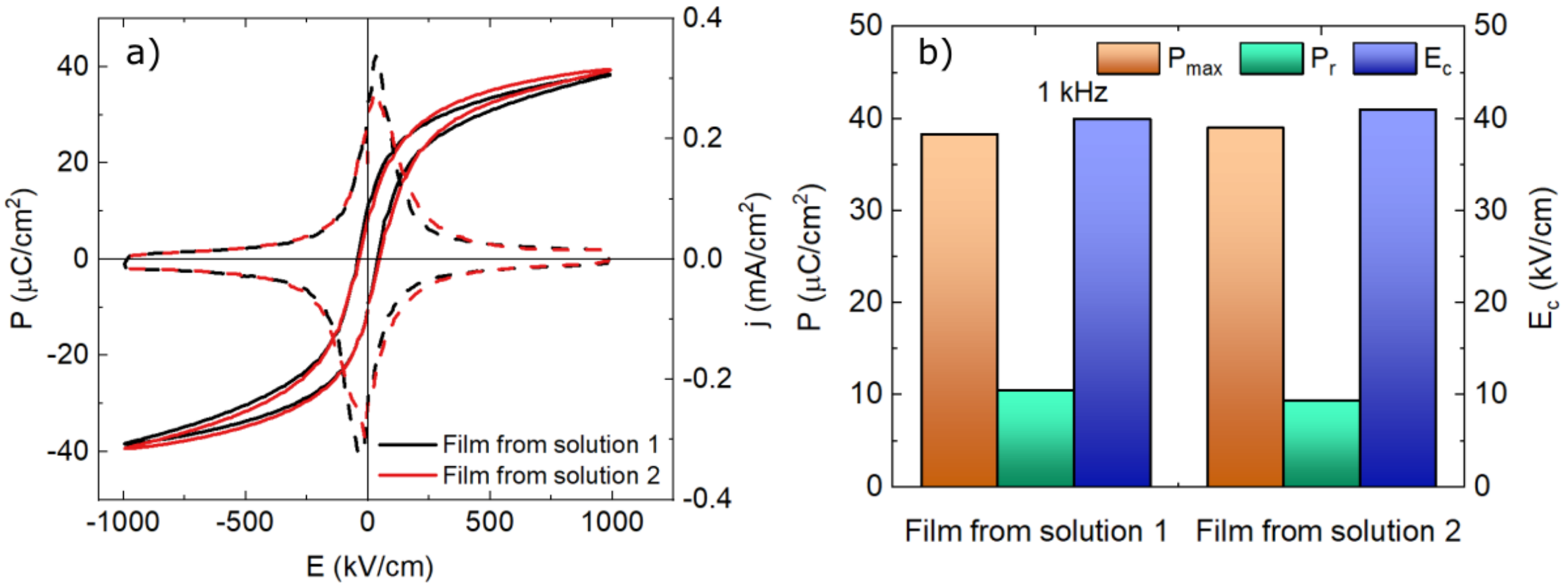

Figure S4: Reproducibility test. **(a)** Polarization *P* and transient current *j* as function of electric field *E* of two different 1 μm thick MN-KNN films, which are prepared from two different batches of precursor solutions. **(b)** Comparison of remnant polarization $P_r$, maximum polarization $P_{max}$, and coercive filed $E_c$ of the two films. Variations of the respective values are 0.7 μC/cm$^2$, 1.1 μC/cm$^2$ and 1 kV/cm.

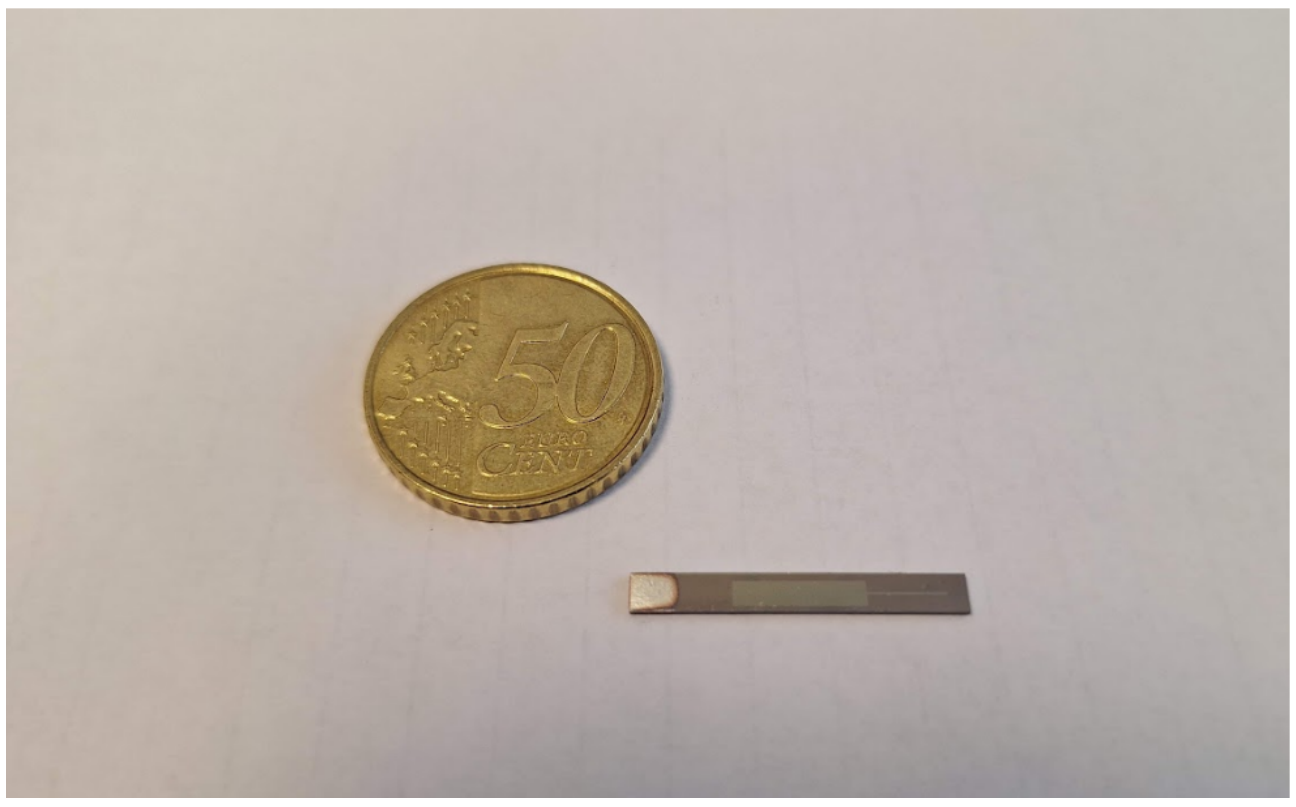

Figure S5: Cantilever used for piezoelectric measurements. The cantilever has a length of 25 mm. A rectangular Pt top electrode is visible at the center, while silver Ag paste at the left edge serves as the bottom contact. A 50-cent euro coin is included for scale.

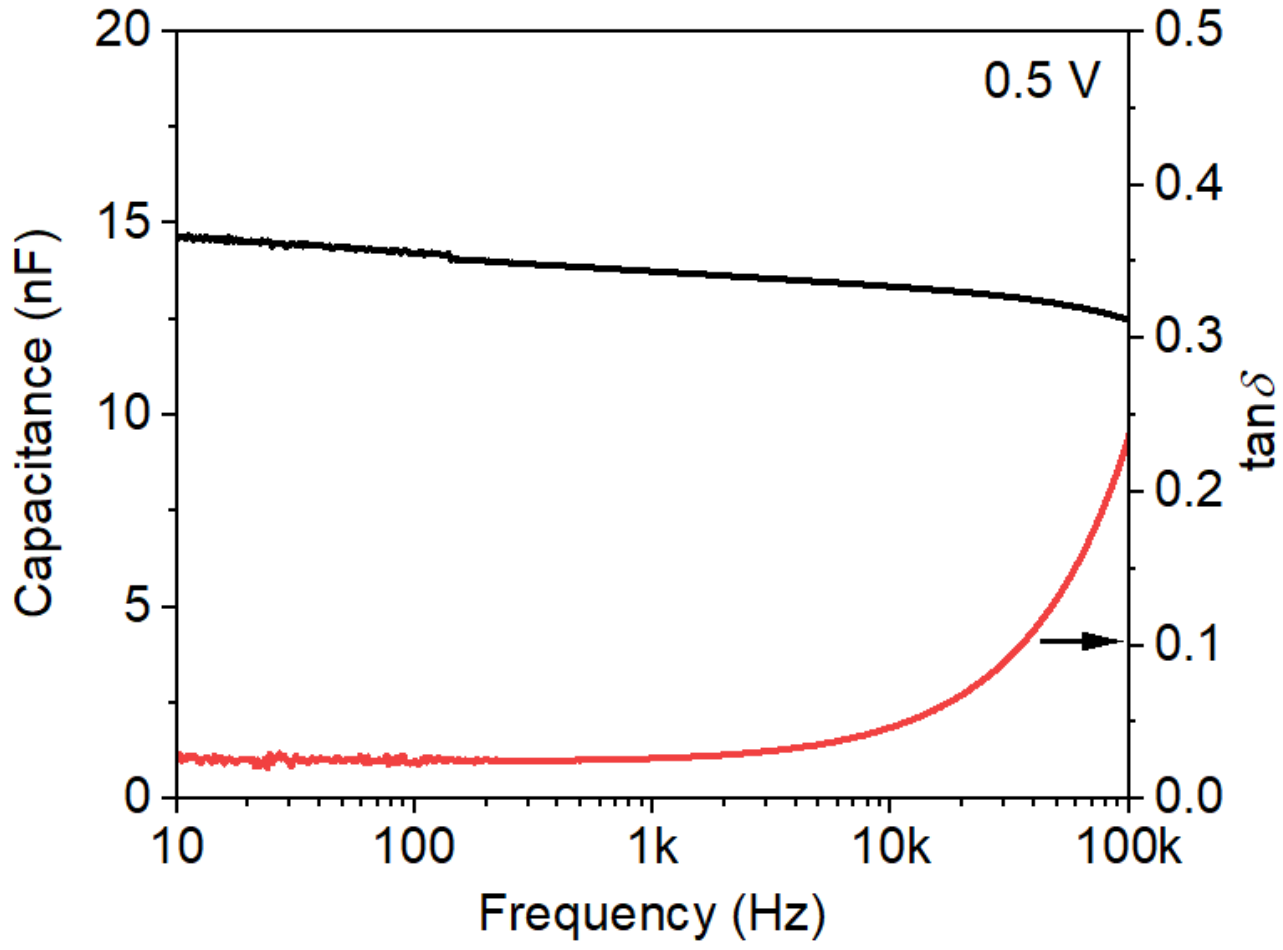

Figure S6: Haptic device. Capacitance $C$ and dielectric losses tan$\delta$ as functions of frequency. The measurement was performed on one of the actuators of the haptic device.

Table S1: Dielectric, ferroelectric, and piezoelectric properties of the lead-free Mn-doped KNN films reported in this work compared to those of the lead-based $Pb(Zr_{0.53}Ti_{0.47})O_3$ (PZT) films on platinized silicon and other lead-free perovskite thin films on various substrates. Abbreviations: Mn-NBT ($Na_{0.5}Bi_{0.5}Ti_{0.995}Mn_{0.005}O_3$); BT-BS ($0.972BaTiO_3$-$0.028BaSnO_3$); NBT-KBT-BT ($0.88Na_{0.5}Bi_{0.5}TiO_3$-$0.08K_{0.5}Bi_{0.5}TiO_3$-$0.04BaTiO_3$); NBT-BT ($0.93Na_{0.5}Bi_{0.5}TiO_3$-$0.07BaTiO_3$); KBT-BMT ($0.95K_{0.5}Bi_{0.5}TiO_3$-$0.05Bi_{0.5}Mg_{0.5}TiO_3$).

| Thin film | Thickness (μm) | $\varepsilon$ | tan$\delta$ | $E_c$ (kV/cm) | $P_r$ (μC/cm$^2$) | $-e_{31,f}$ (C/m$^2$) |
|---|---|---|---|---|---|---|
| PZT [1,2] | 2 | 1620 | 0.034 | 33 | 38 | 17.7 |
| Mn-NBT [3] | 1.12 | - | - | ~75 | 16 | 2.43 |
| BT-BS [4] | 1.2 | 925 | <0.05 | - | - | 1.9 |
| NBT-KBT-BT [5] | 0.35 | 750 | 0.15 | 80-100 | 30 | 2.2 |
| NBT-BT [6] | 2 | 726 | ≤0.05 | ~35 | 13.3 | 12.5 |
| NBT-BT [7] | 3 | ~1300 | ≤0.1 | ~30 | ~16 | 14.4 |
| KBT-BMT [8] | 0.3 | 320 | ≤0.1 | 200 | 60 | 2.2 |
| Mn-KNN (This work) | 1 | 920 | 0.05 | 40 | 10 | 14.8 |

**References**

[1] F. Calame, P. Muralt, Growth and properties of gradient free sol-gel lead zirconate titanate thin films, Appl. Phys. Lett. 90 (2007), 062907.

[2] F. Calame, PZT thin film growth and chemical composition control on flat and novel three-dimensional micromachined structures for mems devices, PhD Thesis (2007). École Polytechnique Fédérale de Lausanne, Switzerland.

[3] B. T. Nguyen, S. S. Won, B. C. Park, Y. J. Jo, C. W. Ahn, I. W. Kim, T. H. Kim, Transverse piezoelectric properties of Mn-doped $Bi_{0.5}Na_{0.5}TiO_3$ thin films, Curr. Appl. Phys. 20 (2020), 1447-1452.

[4] H. Cheng, H. Hida, J. Ouyang, I. Kanno, Electromechanical properties of $BaTiO_3$-$xBaSnO_3$ thin films prepared via combinatorial sputtering, Ceram. Int. 43 (2017), 1597-1601.

[5] M. Abazari, A. Safari, S.S.N. Bharadwaja, and S. Trolier-Mckinstry, Dielectric and piezoelectric properties of lead-free (Bi,Na)$TiO_3$-based thin films, Appl. Phys. Lett. 96 (2010), 082903.

[6] Y. Tanaka, S. Okamoto, K. Hashimoto, R. Takayama, T. Harigai, H. Adachi, and E. Fujii, High electromechanical strain and enhanced temperature characteristics in lead-free (Na,Bi)$TiO_3$–$BaTiO_3$ thin films on Si substrates, Sci. Rep. 8 (2018), 7847.

[7] T. Harigai, Y. Tanaka, H. Adachi, and E. Fujii, Piezoelectric properties of lead-free (Na,Bi)$TiO_3$–$BaTiO_3$ (001) epitaxial thin films around the morphotropic phase boundary, Appl. Phys. Express 3 (2010), 111501.

[8] J. Morishita, H. Kazama, Y. Sato, and H. Funakubo, Epitaxial growth of (Bi,K)$TiO_3$-Bi(Mg,Ti)$O_3$ (001) films and their ferroelectric and piezoelectric properties, Jpn. J. Appl. 58 (2019), SLLB13.